# 3D longitudinal imaging of tumor angiogenesis in mice *in vivo* using Ultrafast Doppler Tomography

Charlie Demené[1,3], Thomas Payen[2], Alexandre Dizeux[2], Guillaume Barrois[2], Jean-Luc Gennisson[1], Lori Bridal[2*], Mickael Tanter[1,3*]

[1]Institut Langevin, ESPCI Paris, CNRS UMR7587, Inserm U979, Paris, France,

[2]Sorbonnne Université, CNRS, INSERM, Laboratoire d'Imagerie Biomédicale, F-75006, Paris, France.

[3] Inserm Accélérateur de Recherche Technologique en Ultrasons biomédicaux, Paris, France

*MT and LB are co-last authors.*

## ABSTRACT

Angiogenesis, the formation of new vessels, is one of the key mechanisms in tumor development and an appealing target for therapy. Non-invasive, high-resolution, high sensitivity, quantitative 3D imaging techniques are required to correctly depict tumor heterogeneous vasculature over time. Ultrafast Doppler was recently introduced and provides an unprecedented combination of resolution, penetration depth and sensitivity without requiring any contrast agents. The technique was further extended to 3D with Ultrafast Doppler Tomography (UFD-T). In this work, UFD-T was applied to the monitoring of tumor angiogenesis *in vivo* providing structural and functional information at different stages of development. UFD-T volume renderings showed that our murine model's vasculature stems from pre-existing vessels and sprouts to perfuse the whole volume as the tumor grows until a critical size is reached. Then, as the network becomes insufficient, the tumor core is no longer irrigated because the vasculature is mainly concentrated in the periphery. In addition to spatial distribution and growth patterns, UFD-T allowed a quantitative analysis of vessel size and length, revealing that the diameter-distribution of vessels remained relatively constant throughout tumor growth. The network is dominated by small vessels at all stages of tumor development with more than 74% of the vessels less than 200 µm in diameter. This study also showed that cumulative vessel length is more closely related to tumor radius than volume, indicating that the vascularization becomes insufficient when a critical mass is reached. UFD-T was also compared with dynamic contrast-enhanced ultrasound (DCE-US) and shown to provide complementary information regarding the link between structure and perfusion. In conclusion, UFD-T is capable of an *in vivo* quantitative assessment of the development of tumor vasculature (vessels with blood speed >1mm/s (sensitivity limit) assessed with a resolution limit of 80 µm) in 3D. The technique has very interesting potential as a tool for treatment monitoring, response assessment and treatment planning for optimal drug efficiency.

## INTRODUCTION

In 1971, Judah Folkman first introduced the idea that tumor growth was angiogenesis-dependent and that this mechanism was a potential target for anti-tumor therapy (Folkman 1971). Indeed, since the oxygen diffusion limit is approximately 250 µm, a tumor cannot exceed a few millimeters in size without a blood supply from its own micro-vascular system providing the essential nutriments and oxygen (Cao and Langer 2008). The initial hypothesis that tumor cells can secrete growth factors to induce sprouting of new blood vessels in their local environment was comforted by numerous studies over the two following decades (Folkman 1990). Remarkably, Folkman envisioned from the beginning, the therapeutic potential of such a mechanism: inhibition of the tumor's blood vessel network would both prevent the increase of the tumor size and potential metastatic dissemination. An exponential number of studies were then conducted to understand the molecular mechanisms of angiogenesis and identify therapeutic targets to fight tumor growth (Weis and Cheresh 2011).

Numerous authors also underlined the heterogeneity and sometimes the inefficiency of the vascular network developed by the tumor (Carmeliet and Jain 2000; Dvorak 1986; Folkman 1995; Weis and Cheresh 2011). Tumor vascular architecture is chaotic and the blood vessels are structurally and functionally anomalous with excessive branching and shunts. From a therapeutic point of view, this poor perfusion reduces efficiency of both chemotherapy and radiotherapy. Many preclinical and clinical trials for antiangiogenic treatments turned out to be disappointing, with the development of a resistance to the treatment or the increased tumor aggressiveness after an initial, slight resorption (Bergers and Hanahan 2008; Ellis and Hicklin 2008). However, another interesting effect was discovered. Under certain conditions, anti-angiogenic treatments could 'normalize' the vascular network of the tumor (Jain 2001), leading to a better physiological condition of the tumor and an increased sensitivity to cytotoxic therapies. New therapeutic strategies were then developed relying on the combined use of anti-antiangiogenics and radiotherapy or chemotherapy (Jain et al. 2011). A very exhaustive review of these strategies has been given by Goel et al (Goel et al. 2011).

Studies have also shown that heterogeneous tumor vascularization both in structure and function can be associated with disease progression and malignancy (Agrawal et al. 2009). Indeed aberrant vascular regions of metabolic insufficiency, ischemia, and necrosis are commonly found in growing tumors

(Gilbertson and Rich 2007). The paradox of reduced vascularization leading to increased tumor growth rate was thoroughly investigated (Carmeliet et al. 1998) and led to the discovery of the Hypoxia-inducible factor-1 (HIF-1), an angiogenesis promoter activated when oxygen levels are substantially reduced (Semenza 2003). The process ultimately results in the selection of the most vigorous, less vascular-dependent tumor cells (Nagy et al. 2010; Yu et al. 2001). These observations imply that tumor vascular heterogeneity may be of interest during diagnosis to anticipate tumor progression in addition to being highly relevant in therapy for efficient response assessment and treatment planning.

In this context, an imaging modality giving both a precise delineation of the tumor vascular bed and information on the local hemodynamics with high spatial and temporal resolution would be very beneficial for preclinical and clinical imaging. The assessment of the structural and functional organization is of the highest interest to study the dynamics of the formation of new vascular networks and to monitor their normalization when anti-angiogenics are administered. Imaging modalities such as µCT or MRI enable 3D imaging with quite high spatial resolution and very good tissue penetration (de Jong et al. 2014). However, their usefulness to evaluate the vascular system remains limited for repeated tumor monitoring due to the need for contrast agent injection and ionizing rays (Kiessling et al. 2004). Furthermore, µCT and MRI do not provide blood flow dynamics, at least not at the single cardiovascular cycle scale. Indirect parameters linked to the time of arrival of the contrast agent can be calculated, but the link with vascular architecture and vessel density is not always obvious (Dighe et al. 2013). Optical techniques, of course, provide better spatial resolutions but with very limited penetration depths, and fields of view are generally limited such that only very small tumors or parts of tumors can be imaged in a reasonable time (Vakoc et al. 2009). A particularly interesting technique is photoacoustic imaging due to the penetration depth of the ultrasound and optical absorption contrasts (Yao and Wang 2014). A non-exhaustive but representative panel of current techniques applied for *in vivo* tumor imaging is given in Table 1.

In this imaging landscape, assessment of tumor angiogenesis via ultrasound is currently contingent on the use of contrast agent (Payen et al. 2015) or very high frequencies (40 to 100MHz) (Ferrara et al. 2000) because conventional ultrasound Doppler techniques suffer from a lack of sensitivity and from the impossibility to extract very slow blood flow (below 10 mm.s$^{-1}$). The recently introduced Ultrafast Doppler technique (Bercoff et al. 2011) can resolve both issues by increasing the sensitivity to blood motion detection by a factor 50 (Macé et al. 2013) compared to conventional techniques and by enabling new paradigms for discrimination between tissue and blood signal using spatiotemporal information (Demené et al. 2015). This gives an unprecedented combination of resolution (in-plane 100 µm at an emission frequency of 15MHz), sensitivity (x50) and detection of slow blood flow (down to 1 mm.s$^{-1}$) for a 2D vascular imaging modality. This has further been extended to 3D with the invention of UltraFast Doppler Tomography (UFD-T), with a first proof of concept made by imaging the rat brain microvasculature (Demené et al. 2016).

In this study, we show that UFD-T is capable of *in vivo* microvascular imaging of tumor development, giving access to high-resolution, quantitative anatomical and functional information in the context of angiogenesis monitoring. In particular, vessel size distribution in the tumor was studied to analyze the development of the vascular tree. Furthermore, by giving a volumic description of the tumor in its entirety, UFD-T overcomes a major pitfall of 2D ultrasound imaging since arbitrary selection of the imaging plane is avoided to provide more relevant and reproducible evaluation of the whole tumor.

## MATERIALS AND METHODS

### 1.1 *Animal protocol*

All animal manipulations have been approved by the French Agriculture Ministry (protocol n° Ce5/2012/082) and the local ethics committee. Mouse Lewis lung carcinoma (3LL) cells (ATCC, Manassas, VA, USA) were injected ($2 \times 10^5$ cells in 100 µL of culture medium) in the flank of a C57BL6 mice (Janvier, Le Genest-St-Isle, France). When the lesion reached approximately 300 mm$^3$, the mouse was euthanized and the tumors was extracted and fragmented. At Day 0, tumor fragments (~10 mm$^3$) were then implanted subcutaneously in the right flank of four, six-week-old female C57BL6 mice (Figure 1). On Days 8, 12, 16 and 20, UFD-T data were acquired on the tumor of each mouse. Mice were placed on a heating pad and flank skin position was maintained using medical adhesive tape to mitigate respiration motion artefacts.

### 1.2 *Ultrafast Doppler Tomography*

UFD-T was performed using an Aixplorer ultrasound scanner (Supersonic Imagine, France) (Figure 1), with custom-dedicated software enabling programmable transmission and reception of plane waves. A high frequency ultrasonic probe (128 elements, 0.08mm pitch, 15MHz central frequency, 85% bandwidth; Vermont, France) was mounted on a custom-motorized set up enabling three degrees of translation (uni-directional repeatability 0.8 µm, bi-directional repeatability ±10 µm; VT-80, Physik Instrumente (PI), Germany) and one degree of rotation (uni-directional repeatability 0.01°, bi-directional repeatability ±0.2°; DT-80, PI). This set-up defined a coordinate system for the mobile probe with a 3-axes $(x', y', z')$ (figure 1) coordinate framework: the x'-axis being the dimension along the piezo elements, the z'-axis beng the axis of ultrasound propagation, and the y'-axis being the out-of-probe-plane direction (Figure 1).It enabled acquisition of UFD images in any zOx imaging plane. Acoustic contact was provided by a large amount of acoustic gel on and around the tumor beneath a water-filled acoustic tank within which the ultrasonic probe could move freely. To reconstruct a 3D volume with a good isotropic resolution despite the natural poor elevation focusing capability of the ultrasonic probe, the strategy is to scan the whole volume with different angular probe orientations (Demené et al. 2016). For a given orientation θ (see Figure 1, θ is the angle $\widehat{xOx'}$ ) of the probe, a volume is acquired slice by slice by translating the probe every 200 µm along y'. Then the probe is rotated along θ with 10° steps, giving a total of 18 scan volumes that are then merged in a post-processing step. Depending on the size of the region of interest, the number of translations is adapted. A typical 52-step scan provides a 10.4-mm-wide square in the horizontal (xOy) orientation. Depth is only limited by absorption of ultrasound and can be up to 25 mm with 15 MHz transmit frequency. The scan volume was 10.4x10.4x(12 to 18) mm because this is the maximum volume requiring only one scan for each $\theta$ direction. Because the probe is 128 elements x0.08 mm = 10.24 mm wide, the scan along the y-direction cannot be longer than 10.24 mm without losing the overlap between the scanned volumes. With a 0.2mm scan step, we chose to stop the y' scanning when 10.4 mm was reached. For each slice (a plane position parallel to x'0y'), UFD data consists of 400 ultrasonic frames (each frame being made of 8 compounded, in phase quadrature beamformed ultrasonic images obtained via plane-wave-emissions with different angles (angles -7° to 7° with a 2° step relative to horizontal) (Montaldo et al. 2009)), are acquired at a frame rate of 500 Hz (the pulse repetition frequency PRF is then 4000 Hz). Respiratory and cardiac gating are performed on each acquisition to avoid respiratory artefacts and to ensure that three synchronized cardiac cycles are included in the analyzed data. Each UFD dataset is then filtered using a dedicated spatiotemporal Singular Value Decomposition filter whose implementation is decribed in Demené et al. (2015), and the energy is computed in each pixel to obtain a classical UFD Power image. The complete data set consists of 18 UFD Power volumes (one volume representing 52 slices) which are co-registered using correlation techniques and summed. The 3D Point Spread Function (PSF) of this entire imaging scheme is simulated using Field II software (Jensen 2010). To sum up, a linear array with acoustic caracteristics similar to the one used in the experiment has been designed for simulation in Field II (128 elements, 0.08mm pitch, 15MHz central frequency, 85% bandwidth, elevation width 1.1 mm, elevation focus 5 mm). This linear array was then used to scan across the elevation direction for a point source placed at the focus, with simulation of RF Data for the 8 angles of plane wave emission described previously. The resulting RF data were beamformed with the same beamformer as applied to the real data (delay and sum beamformer, f/d = 1) and the image obtained was displayed in terms of the power (square of the amplitude). Then the longitudinal scans were rotated using 18 angles (10° steps) to emulate the real scanning process, and summed to generate the $psf(x, y, z)$ of our system. This PSF was then used as a deconvolution kernel in a 3D Wiener filter enabling recovery of the resolution lost during the averaging process (back to 80 µm). In brief, a Wiener filter $\widehat{W}(k_x, k_y, k_z)$ is designed in k-space (being the reciprocal space of the fixed coordinate system $(x, y, z)$ of figure 1) based on the input signal Fourier energy spectrum $S(k_x, k_y, k_z)$, the variance of the noise $\sigma^2$ and $PSF(k_x, k_y, k_z)$ the Fourier transform of $psf(x, y, z)$ obtained in the previously described simulation process.

$$\widehat{W}(k_x, k_y, k_z) = \frac{1}{PSF(k_x, k_y, k_z)} \cdot \frac{S(k_x, k_y, k_z) - \sigma^2}{S(k_x, k_y, k_z)} \qquad (1)$$

The filtered UFD-T volume $\hat{e}(x, y, z)$ was ultimately computed as follows:

$$\hat{e}(x, y, z) = \text{TF}^{-1}[\widehat{W}(k_x, k_y, k_z) \cdot S(k_x, k_y, k_z)] \qquad (2)$$

A complete and detailed description of the implementation of this 3D reconstruction and filtering process can be found in (Demené et al. 2016).

The spatial variation of the PSF is not taken into account, and cannot be using a Wiener filtering formalism since it is based on convolution. We can, therefore, expect better quality around the elevation focus. However it should be noted that the shape of the 3D PSF is mainly driven by the scanning process, which is similar across space and only marginally driven by the spatial variations of the in-plane ultrasonic PSF of the beamforming process.

The reconstructed volumes were then rendered and processed using Amira® software (Thermo Fisher Scientific, Waltham, Massachusetts, USA). Amira® is a data visualization software enabling volume rendering and proposing 3D algorithms and quantification tools. Length and diameter of vessels across the vascular network of each tumor were calculated using the skeletonisation tool. For such an algorithm to work, a minimum signal threshold has to be defined for a voxel to be considered to be within a vessel. This level was set as a z-score ($>3\sigma$, $\sigma$ being the standard deviation of spatial noise). The skeletonisation typically performs a medial axis transform to return the center line of each vessel (Lee 1982). The skeletonisation then returns the list of all vessel segments with their lengths and mean diameters. This population of vessels can then be gathered into 0.04mm-wide diameter bins to calculate the total length of sub-populations of vessels with a particular average diameter (Figure 5 and Figure 6). *Angiogenesis* being the physiological process through which new blood vessels form from pre-existing vessels, a quantification of angiogenesis needs to put a number on this vasculature, for example the partial volume it represents in the tumor. Blood vessels being in first approximation cylindrical, representing their length according to their diameter gives both an indication of that volume and of the repartition of blood vessels between big and small vessels.

### 1.3 *Dynamic Contrast-Enhanced Ultrasound (DCE-US)*

Contrast imaging was performed after the UFD-T experiments to avoid MB interference in the UFD-T results. Sonovue™ microbubbles (Bracco Suisse, Geneva, Switzerland) were used for the assessment of vascularization (Figure 1). Before use, the contrast agents were reconstituted in 0.9% saline solution, resulting in a microbubble (MB) suspension containing $2 \times 10^9$ MBs/mL. A Sequoia 512 US system (Acuson, Siemens, Mountain View, CA, USA) with the 15L8w probe was used to perform contrast imaging at 12

MHz along two planes close to the tumor's maximal cross section and separated by 1 mm. The non-linear signals from the microbubbles were specifically detected with cadence contrast pulse sequencing (CPS). MB were injected as a bolus in the tail vein at 1mL/kg using a controlled system described in a previous publication (Dizeux et al. 2016) and a flow of 2 mL/min. Two perfusion parameters were then extracted from 60-s recordings using parameters assessed without model fitting (Barrois et al. 2013). The Time-Of-Arrival (TOA) was calculated in 3 by 3 pixel regions based on the time point at which contrast agent intensity reached 90% of its maximum intensity in the region. If intensity remains below 3 times the noise standard deviation in a region, then the region is considered a non-perfused area. The Mean of Intensity (MoI) is the temporal average of the low-pass filtered intensity. Here, the MoI and TOA averaged over all pixels in all maps are reported. To describe the distribution of these parameters in each map, the average standard deviations for all measurements are also reported. The individual maps of the parameters were used to spatially compare DCE-US and tomography results.

### 1.4 *Statistical analysis*

Statistics on quantitative results were performed using Matlab. A global one-way ANOVA test was used to evaluate if the samples are from different populations with the same mean. A t-test was then performed to assess the difference between the results obtained in two groups. A p-value < 0.05 was considered to indicate a significant difference.

## RESULTS

UFD-T was performed on all tumors to monitor their structural and functional development. Tumors were identified and segmented manually on the successive B Mode images acquired during UFD-T. From the same acquisitions, UFD-T reconstruction enabled the detection of blood vessels, consequently there was no need to co-register the tumor boundaries and the vascular network which were then fused directly for volume rendering (Figure 2a).

The noninvasiveness and ease-of-use of the technique allowed longitudinal imaging of the tumor development across our 4 mice at Days 8, 12, 16 and 20 (Figure 2b**Erreur ! Source du renvoi introuvable.**). However, 2 mice had to be euthanized before the end of the study as they met the endpoint criteria regarding tumor size. One mouse was also not imaged at Day 16 due to technical difficulties. Steady tumor growth was observed with tumor volume increasing from 11 ± 3.3 mm$^3$ at Day 8 to 473 ± 188 mm$^3$ at Day 20 (68 ± 15 mm$^3$ at Day 12 and 194 ± 104 mm$^3$ at Day 16) as seen on Figure 2b. In addition, the 3D renderings clearly showed that the vasculature had a tendency to grow on the periphery rather than in the tumor core. The apparent smaller density of vessels of the caliber detected by ultrafast Doppler in the center of the tumor is consistent with the vascular architecture that has been previously reported for this type of tumor (3LL)(Dizeux et al. 2017).

UFD-T 3D monitoring revealed the process involved in building the vascular architecture irrigating the tumor. These results demonstrated that, in our model, the tumor's blood supply stemmed from pre-existing blood vessels (Figure 3, Supplemental Video 1). Blood vessels were observed as a loose network around the early-stage tumor and did not appear to penetrate the tumor indicating that it was still mostly the normal vascular network already present in the skin. However, as the tumor volume increased, exceeding a few millimeters in diameter, sprouting of new vessels from existing surrounding blood vessels was seen. Between Day 12 and Day 16, the vascularization kept on growing developing into a more homogeneous network irrigating the whole tumor volume. Then, as a critical size was reached, a vascular void was observed in the center of the tumor surrounded by highly-perfused periphery with a dense and tortuous network (Supplemental Video 1). The precise delineation of vessels provided by UFD-T also made it possible to follow individual vessels in time. For example, a large pre-existing skin vessel was identified from one time point to the next as highlighted with a green line on Figure 3. From this particular vessel, it can be observed that at each stage new blood vessels are sprouting and forming ramifications leading to a global increase of the vascular network.

The DCE-US results are reported in Table 2. The MoI and TOA averaged over all pixels are reported as a "spatial mean". Distribution of these parameters in each map is reflected in the average standard deviations for all measurements reported as "spatial SD".

The four mice tend to follow similar trends. MoI increases from Day 8 to Day 12 while TOA decreases as the vascular networks grow. Then, as the tumor develops, these trends reverse. A particularly interesting aspect is the increase in standard deviations at later stages for both parameters reflecting the higher spatial heterogeneity of the vasculature.

The evolution of the vascular network supplying the tumor seen on UFD-T volume renderings was also confirmed by parametric maps acquired at the same time points using DCE-US (Figure 4). Time of arrival (TOA) and Mean of Intensity (MoI) maps clearly depict the different steps of tumor vasculature development. Whereas MoI informs on perfusion density, TOA reflects the spatial hierarchy in terms of blood supply. Early on, the network is scattered and poorly ramified resulting in high TOA values with a dispersed distribution and low MoI except in the few pre-existing big vessels. Then, the vascular network grows leading to a decrease in TOA across the tumor. Higher overall MoI values are observed as well which is also in line with a more efficient, developing network. The tumor then reaches a critical size, and the blood vessels, mainly located at the periphery of the tumor, do not supply the center anymore, resulting in TOA and MoI maps with radial gradients. In the TOA map shown in Figure 4, two pixels located in the center of the tumor are displayed in black. They are considered as non-perfused because the intensity did not reach 3 times the noise standard deviation. In every case assessed in this study, the inhomogeneity seen on DCE-US maps was also observed in the distribution of blood vessels as assessed by UFD-T.

Beyond the visual anatomic descriptive capabilities of UFD-T for tumor imaging, this new modality also enables quantitative assessment of the state of the tumor. In this paper we chose to explore the quantification capabilities of UFD-T by computing relevant anatomical parameters such as the length of vessels irrigating the tumor, their diameter and the tumor volume. We first quantified the growth of the vascular network of the tumor by calculating the length of blood vessel at the different stages of development (Figure 5). For data analysis, the results obtained at Day 16 and Day 20 were combined to get 4 measurements at each growth stage. In order to take into account that not all blood vessels are equivalent, we sorted them according to their diameter and computed the cumulative length per population of vessels gathered in 0.04mm-wide diameter bins (Figure 5.a). First, the total vascular length, calculated as the sum of all the bins, increases as the tumor grows showing a quick evolution from Day 8 to Day 12 (208.6 ± 51.6 mm to 424.7 ± 99.6 mm, $p < 0.01$), and a slower pace from Day 12 to later time points (548.0 ± 156.1 mm, not significantly different). Interestingly, this inter-stage increase is seen across all diameters. For each stage of development, the repartition of the blood vessels according to their diameter follows an exponential tendency ($R^2 > 0.93$) that has been seen in previous studies (Hashizume et al. 2000; Junaid et al. 2017). The degree of growth is quantified finely according to the hierarchical position of the vessel in the vascular architecture: a large part of the network is represented in small vessels less than 200 μm in diameter (80% at Day 8, 79% at Day 12, and 74% at later stages).

In addition, UFD-T results allowed us to quantify whether, beyond the obvious growth of the vascular network of the tumor, there is a structural change in the distribution of vessels in the vascular tree: does the tumor growth result in more big or small vessels? For this purpose, the cumulative length is represented not in millimeters but as a percentage of the total length of the network of each tumor (Figure 5.b). The average value of this "distribution function" of the vessels length is shown by grouping the 4 time points of the study. The specific results for each development stage are also displayed. These results indicate that, despite a strong growth of the tumor vascular network, the overall structure and ratio of the small vessels relative to the larger vessels remains generally the same during tumor growth. However, a slight difference can be observed. The relative length of vessels at Day 8 was higher than at later stages for vessels smaller than 160 μm (although not significantly) and lower between 160 and 250 μm ($p = 0.02$).

Finally, the efficiency of the vascular network was investigated by correlating the cumulative length of vessels $\zeta$ to the tumor volume and radius (Figure 6). It is reasonable to assume that in a given healthy tissue there is a relationship of proportionality between the volume of tissue considered and the length of the vascular network that irrigates it. Here, however, this is not the case and $\zeta$ rather changes according to the radius of the tumor rather than to the volume. Thus, Figure 6a shows that $\zeta$ normalized by the tumor volume dramatically decreases with the age of the tumor, which suggests a weakening of the tumor perfusion, whereas in the Figure 6.b, $\zeta$ normalized by the tumor radius exhibits quite similar trends across the 3 stages of development.

## DISCUSSION

UFD-T is a novel imaging technique for quantitative 3D visualization of vasculature. In this study, 4 ectopic tumors were imaged longitudinally over 13 days to provide various information on the development of their vascular tree. Three-dimensional UFD-T results showed the structural and functional organization of the growing tumors revealing not only a steady volume increase but also a tendency of the network to develop primarily in the periphery for this particular type of tumors (3LL)(Dizeux et al. 2017). High-resolution individual vessel monitoring showed that the tumor vasculature stems from pre-existing vessels. The quality of vascular structure reconstruction can be regarded as slightly inferior to that obtained in a previous study on the rat brain (Demené et al. 2016). This can be explained both because the vascular network of the tumor is intrinsically less organized than the rat brain vasculature, but also because motion artefacts are more important. Smaller vessels sprout as the tumor grows, irrigating the full tumor volume until a critical size is reached. The tumor core then generally lacks sufficient perfusion and ultimately becomes necrotic. Further analysis on vessel diameter and length confirmed the tumor functional evolution providing additional quantitative information on the vascular tree. As expected, the network increases over time with a large part constituted by small vessels at all stages. However, the vessel diameter distribution remains relatively constant throughout tumor growth in our model. This study also showed that vessel length seems to be more related to tumor radius than to volume leading to the perfusion becoming insufficient when critical mass is reached. UFD-T allowed a precise 3D description of the growth of the tumorous vascular tree through angiogenesis from preexisting vessels to a chaotic and weakly efficient network leading to areas of necrosis due to lack of nutriments and oxygenation.

In current ultrasound imaging, parameters are typically evaluated in 2D imaging planes, generally as spatial means. However, solid tumor vascular distribution is often very heterogeneous both spatially and in terms of vessel diameter and flow rates (Gillies et al. 1999; Jain 1988). The precise characterization of tumor vasculature thus requires a 3D imaging technique providing both structural and functional information that is sensitive enough to assess vessels as small as a few hundred microns in diameter and flows as slow as 1 mm/s. In addition, 3D imaging is particularly important for longitudinal monitoring as it removes the variability in 2D imaging plane selection from one time point to another allowing single vessel analysis as shown in this work. Finally, functional imaging needs to be coupled with a reliable, co-registered imaging technique such as B Mode imaging to identify where the vascular features are located. This study showed that UFD-T fits all the criteria to be a very useful tool for the monitoring of tumor vascular development through angiogenesis. In the future, the development of high frequency 3D probes and the electronics to drive them at ultrafast framerates will enable acquisition of volumic data without the need for time-consuming mechanical scanning.

When looking at the relative distribution of the vessel diameter, a slight difference was observed with respect to the relative length of vessels at Day 8 which was higher than at later stages for vessels below 160 µm and lower between 160 and 250 µm. It may be speculated that inflammation of the tumor occurring at later stages may have led to dilatation of very small vessels reflected by this translation towards the vessels of larger diameter. In further works, histology would be necessary to test this hypothesis and understand why only this range of vessels is concerned.

Studies have shown that the vascular heterogeneity of tumors can be linked to disease progression and malignancy (Agrawal et al. 2009). Beyond a certain size, simple diffusion of oxygen to active tumor cells becomes inadequate, and the tumor needs to turn toward another way to sustain its growth. In the early 1990s, studies showed that hypoxia could induce expression of growth factors in tissue culture (Kourembanas et al. 1990; Shweiki et al. 1992). Investigation of the mechanism behind hypoxia-stimulated angiogenesis revealed the role of the Hypoxia-inducible factor-1 (HIF-1), a transcription factor responsible for cellular adaptive responses to hypoxia (Pugh and Ratcliffe 2003). Under hypoxic conditions, HIF-1 induces the expression of various factors such as VEGF, inducing angiogenesis and protecting endothelial cells. In addition, HIF-1-regulated genes have been associated with resistance to chemotherapy, invasion, metastasis, and tumor malignant phenotypes. To date, extensive efforts have been devoted to understand the role of HIF-1 and hypoxia in tumor growth and resilience to treatment (Carmeliet et al. 1998). In addition, Harada et al. (Harada et al. 2008) showed that the distance between HIF-1-active regions and the nearest tumor blood vessel correlates to the diameter of the vessel. In this context, UFD-T could prove very useful to quantify vessel diameters and map their 3D distribution which could help to identify HIF-1-active regions, prone to angiogenesis.

UFD-T would also greatly benefit therapy for efficiency evaluation, response assessment and treatment planning. For example, anti-angiogenic therapy could profit greatly from UFD-T not only to monitor the early effects on vascularization but also to improve treatment efficiency. Certain antiangiogenic agents have been seen to normalize the chaotic tumor vasculature increasing the efficacy of complementary therapies such as chemotherapy (Jain 2005). However, a careful schedule needs to be respected as the normalization is transitory. Imaging techniques such as UFD-T may identify this optimal window as well as provide information on the underlying mechanisms of vasculature normalization. The technique would thus be a crucial tool not only for cancer but also other diseases with abnormal vasculature such as psoriatic skin lesions and atherosclerotic plaques.

Contrast imaging was performed in this work to validate UFD-T observations. DCE-US parameters confirmed the evolving characteristics of tumor development seen with UFD-T. TOA and MoI maps were calculated providing information on perfusion from the scattered sprouts stemming from a pre-existing vessel to the heterogeneous network with a weakly perfused central region reflected by a clear radial gradient on both DCE-US maps. However, some limits apply to the comparison resulting in slight differences between the results obtained with the two modalities. First of all, DCE-US is a 2D technique as it is currently performed. In this work, the corresponding plane in UFD-T was carefully found in the 3D volume based on the B-mode images acquired during both scans, but some variability remains possible. Second, as mentioned before, UFD-T spatial resolution is 100 µm with our parameters (emission frequency of 15 MHz) and it can detect flows as slow as 1 mm/s. However, a minimum number of red blood cells is required to get a signal. With DCE-US, the spatial resolution is lower (on the order of 200 µm), but the technique can theoretically detect presence of very slow flows in vessels as small as a microbubble (1-3 µm in diameter) due to its sensitivity at the level of individual microbubbles. UFD-T can distinguish vessels that are very close to one another and precisely delineate the vasculature whereas DCE-US has the edge when it comes to very weak flows and smaller vessel detection such as tumor capillaries whose diameter is on the order of a few microns (Leunig et al. 1992). Thus, UFD-T vascular visualization can be completed using DCE-US detection of very poorly perfused areas. Finally, even though 3D spatial resolution has been improved with UFD-T down to ~100µm, tumor vessels can be as small as a few µm with quasi-static flows. It may be possible that their diameter was overestimated and they were sorted in the 100µm bin in the graphs and for analysis.

At the time this article was being written, 2D super resolution ultrasound using contrast agents as isolated ultrasound scatterers were starting to be used for 2D super-localised vascular imaging of tumors (Opacic et al. 2018) with very interesting quantification capabilities. The differences between the two studies has to be pointed out because they reveal two different approaches. We aimed at giving a 3D-capable imaging technique that can achieve completion in a reasonable time with relevant quantitative data for angiogenesis monitoring. The time to achieve the complete 3D scan (with ~65 slices per volume and 18 orientations) is typically 15 to 20 minutes.

With the technique described in the cited paper (Opacic et al), taking into the account the 40s per super resolved image acquisition time, building an equivalent 3D volume (with the same number of slices and orientation, which is a gross estimation of what would be needed) would take 13h. Therefore, contrast agent and UFD-T may ultimately be combined to perform super-resolution vascular imaging by using both increased concentration of micro-bubbles (which needs to remain high and relatively stable, using infusion of microbubbles with a syringe pump) and plane wave ultrafast imaging (Errico et al. 2015), in order to reduce the acquisition time. Two important issues are that super-resolution imaging is very sensitive to tissue motion which needs to be perfectly corrected as recently demonstrated (Foiret et al. 2017; Hingot et al. 2017), and also that 3D fusion of 2D super-resolved images is still an open question because the out-of-plane dimension is not being super-resolved.

This pilot study evaluates the potential of UFD-T in tumor development and anti-angiogenic monitoring. A complementary study in histology would be necessary to more thoroughly discuss the relevance of the curves of vessel length obtained from the UFD-T data: it is possible in microscopy to count the vessels and their relative proportion as a function of the diameter. Hashimuze et al. (Hashizume et al. 2000) investigated the cause for the well-known leakiness of tumor vessels. Using fluorescence staining, immunohistochemistry as well as Scanning Electron Microscopy, they studied the characteristics of individual vessels including their diameter distribution. Interestingly, the curve obtained strongly resembles the power law observed here non-invasively with UFD-T (Figure 5). Additional structural parameters could also be accessed using UFD-T such as tortuosity and the degree of ramification. Tortuosity has been particularly investigated, showing a positive correlation to tumor development, and it is one of the markers of vessel normalization resulting from antiangiogenic treatment (Gillies et al. 1999).

Beyond these structural markers, however, UFD-T can also give access to other perfusion parameters such as the axial blood flow speed in every pixel of a 3D UFD-T acquisition. Although this parameter was not calculated in this study due to the trade-off in framerate it requires, previous studies (Demené et al. 2016) have shown that UFD-T can discriminate flows based on their speed using a bank of band-pass filters. As the orientation of the vessel is known in the 3-D space, it is even possible to estimate the true speed vector discriminating veins from arteries although diverging errors will apply for horizontal vessels.

## CONCLUSION

UFD-T is a non-invasive 3D imaging technique capable of quantifying tissue vascularization *in vivo*. This study showed that UFD-T can give access to structural and functional information in tumors during a longitudinal follow-up revealing the specific stages of angiogenesis. The technique revealed that our model's vasculature stems from pre-existing vessels and then sprouts quickly to perfuse the whole volume. When the tumor reaches a critical size, the core is no longer irrigated ultimately resulting in a necrotic core. UFD-T high-resolution allows the extraction of quantitative parameters such as the vessel diameter revealing a relatively constant size distribution throughout tumor growth in our model. UFD-T also showed that the vascular tree grew as a function of tumor radius rather than volume explaining the appearance of the necrotic central region. Heterogeneity is a typical trademark of tumor vascular development through angiogenesis leading to the necessity of a 3D technique to evaluate this aspect crucial to therapy for treatment monitoring, response assessment and treatment planning for optimal efficiency.

# TABLES

| Modality | MRI | CT (VCT) | μCT | Photoacoustics | OCT(OFDI) |
|---|---|---|---|---|---|
| Resolution | ~100-500 μm | ~50-200 μm | ~30-200 μm | ~50-100 μm | 10 μm |
| Penetration Depth | No limit | No limit | No limit | 1 cm | 1 mm |
| Acquisition time | A few minutes to hours | Tens of minutes | Tens of minutes | Tens of minutes | Tens of minutes |
| Need for contrast agent | yes | Yes | yes | no | no |
| Example | 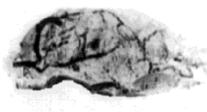 | 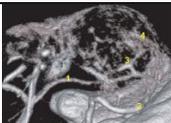 | 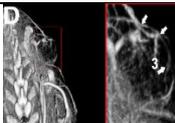 | 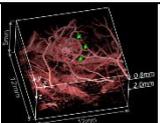 | 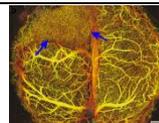 |
| Reference | (Kobayashi et al. 2001) | (Kiessling et al. 2004) | (Ehling et al. 2014) | (Laufer et al. 2012) | (Vakoc et al. 2009) |

**Table 1** *Review of some of the tumor vascular imaging techniques in preclinical research in vivo.*

|  | Spatial mean | | Spatial SD | |
|---|---|---|---|---|
| Day | MoI [a.u.] | TOA [s] | MoI [a.u.] | TOA [s] |
| 8 | 79.7 ± 5.1 | 28.6 ± 8.7 | 2.2 ± 1.6 | 8.3 ± 3.7 |
| 12 | 84.2 ± 3.7 | 19.0 ± 7.5 | 1.7 ± 1.0 | 8.0 ± 3.3 |
| 16-20 | 75.6 ± 3.1 | 33.7 ± 10.0 | 7.1 ± 0.8 | 22.9 ± 11.0 |

**Table 2** *DCE-US results for the 4 mice. The mean and Standard Deviation (SD) of the Mean of Intensity (MoI) and Time-Of-Arrival (TOA) are shown for the different stages.*

## ACKNOWLEDGMENTS

The research leading to these results has received funding from the European Research Council under the European Union's Seventh Framework Programme (*FP7/2007-2013*) / ERC *grant agreement* n° 339244-FUSIMAGINE. We also thank the CEF and PIV of the LIB for their help with the imaging platform.

# REFERENCES

Agrawal G, Su M-Y, Nalcioglu O, Feig SA, Chen J-H. Significance of breast lesion descriptors in the ACR BI-RADS MRI lexicon. Cancer 2009;115:1363–80.

Barrois G, Coron A, Payen T, Dizeux A, Bridal SL. A multiplicative model for improving microvascular flow estimation in dynamic contrast-enhanced ultrasound (DCE-US): theory and experimental validation. IEEE Trans Ultrason Ferroelectr Freq Control 2013;60:2284–2294.

Bercoff J, Montaldo G, Loupas T, Savery D, Mézière F, Fink M, Tanter M. Ultrafast compound Doppler imaging: providing full blood flow characterization. IEEE Trans Ultrason Ferroelectr Freq Control 2011;58:134–47.

Bergers G, Hanahan D. Modes of resistance to anti-angiogenic therapy. Nat Rev Cancer 2008;8:592–603.

Cao Y, Langer R. A review of Judah Folkman's remarkable achievements in biomedicine. Proc Natl Acad Sci 2008;105:13203–13205.

Carmeliet P, Dor Y, Herbert J-M, Fukumura D, Brusselmans K, Dewerchin M, Neeman M, Bono F, Abramovitch R, Maxwell P, Koch CJ, Ratcliffe P, Moons L, Jain RK, Collen D, Keshert E. Role of HIF-1A inhypoxia- mediated apoptosis, cell proliferation and tumour angiogenesis. Nature 1998;394:485–490.

Carmeliet P, Jain RK. Angiogenesis in cancer and other diseases. Nature 2000;249–257.

de Jong M, Essers J, van Weerden WM. Imaging preclinical tumour models: improving translational power. Nat Rev Cancer Nature Publishing Group, 2014;14:481–493.

Demené C, Deffieux T, Pernot M, Osmanski BF, Biran V, Gennisson JL, Sieu LA, Bergel A, Franqui S, Correas JM, Cohen I, Baud O, Tanter M. Spatiotemporal Clutter Filtering of Ultrafast Ultrasound Data Highly Increases Doppler and fUltrasound Sensitivity. IEEE Trans Med Imaging 2015;34:2271–2285.

Demené C, Tiran E, Sieu LA, Bergel A, Gennisson JL, Pernot M, Deffieux T, Cohen I, Tanter M. 4D microvascular imaging based on ultrafast Doppler tomography. NeuroImage 2016;127:472–483.

Dighe S, Blake H, Jeyadevan N, Castellano I, Koh D-M, Orton M, Chandler I, Swift I, Brown G. Perfusion CT vascular parameters do not correlate with immunohistochemically derived microvessel density count in colorectal tumors. Radiology 2013;268:400–10.

Dizeux A, Payen T, Barrois G, Le Guillou Buffello D, Bridal SL. Reproducibility of Contrast-Enhanced Ultrasound in Mice with Controlled Injection. Mol Imaging Biol Molecular Imaging and Biology, 2016;18:651–658.

Dizeux A, Payen T, Le Guillou-Buffello D, Comperat E, Gennisson J-L, Tanter M, Oelze M, Bridal SL. In Vivo Multiparametric Ultrasound Imaging of Structural and Functional Tumor Modifications during Therapy. Ultrasound Med Biol 2017;43:2000–2012.

Dvorak HF. Tumors: wounds that do not heal. Similarities between tumor stroma generation and wound healing. N Engl J Med 1986;315:1650–9.

Ehling J, Theek B, Gremse F, Baetke S, Möckel D, Maynard J, Ricketts SA, Grüll H, Neeman M, Knuechel R, Lederle W, Kiessling F, Lammers T. Micro-CT imaging of tumor angiogenesis: Quantitative measures describing micromorphology and vascularization. Am J Pathol American Society for Investigative Pathology, 2014;184:431–441.

Ellis LM, Hicklin DJ. Pathways Mediating Resistance to Vascular Endothelial Growth Factor–Targeted Therapy. Clin Cancer Res 2008;14:6371–6375.

Errico C, Pierre J, Pezet S, Desailly Y, Lenkei Z, Couture O, Tanter M. Ultrafast ultrasound localization microscopy for deep super-resolution vascular imaging. Nature Nature Publishing Group, 2015;527:499–502.

Ferrara KW, Merritt CRB, Burns PN, Stuart Foster F, Mattrey RF, Wickline SA. Evaluation of tumor angiogenesis with US: Imaging, Doppler, and contrast agents. Acad Radiol 2000;7:824–839.

Foiret J, Zhang H, Ilovitsh T, Mahakian L, Tam S, Ferrara KW. Ultrasound localization microscopy to image and assess microvasculature in a rat kidney. Sci Rep 2017;7:13662.

Folkman J. Tumor angiogenesis: therapeutic implications. N Engl J Med 1971;285:1182–6.

Folkman J. What Is the Evidence That Tumors Are Angiogenesis Dependent? J Natl Cancer Inst 1990;82:4–7.

Folkman J. Angiogenesis in cancer, vascular, rheumatoid and other disease. Nat Med 1995;1:27–31.

Gilbertson RJ, Rich JN. Making a tumour's bed: glioblastoma stem cells and the vascular niche. Nat Rev Cancer 2007;7:733–736.

Gillies RJ, Schornack PA, Secomb TW, Raghunand N. Causes and effects of heterogeneous perfusion in tumors. Neoplasia 1999;1:197–207.

Goel S, Duda DG, Xu L, Munn LL, Boucher Y, Fukumura D, Jain RK. Normalization of the Vasculature for Treatment of Cancer and Other Diseases. Physiol Rev 2011;91:1071–1121.

Harada H, Xie X, Itasaka S, Zeng L, Zhu Y, Morinibu A, Shinomiya K, Hiraoka M. Diameter of tumor blood vessels is a good parameter to estimate HIF-1-active regions in solid tumors. Biochem Biophys Res Commun 2008;373:533–538.

Hashizume H, Baluk P, Morikawa S, McLean JW, Thurston G, Roberge S, Jain RK, McDonald DM. Openings between defective endothelial cells explain tumor vessel leakiness. Am J Pathol American Society for Investigative Pathology, 2000;156:1363–80.

Hingot V, Errico C, Tanter M, Couture O. Subwavelength motion-correction for ultrafast ultrasound localization microscopy. Ultrasonics 2017;77:17–21.

Jain R, Gutierrez J, Narang J, Scarpace L, Schultz LR, Lemke N, Patel SC, Mikkelsen T, Rock JP. In vivo correlation of tumor blood volume and permeability with histologic and molecular angiogenic markers in gliomas. Am J Neuroradiol 2011;32:388–394.

Jain RK. Determinants of tumor blood flow: a review. Cancer Res 1988;48:2641–2658.

Jain RK. Normalizing tumor vasculature with anti-angiogenic therapy: a new paradigm for combination therapy. Nat Med 2001;7:987–9.

Jain RK. Normalization of tumor vasculature: an emerging concept in antiangiogenic therapy. Science 2005;307:58–62.

Jensen JA. Users' guide for the Field II program. 2010.

Junaid TO, Bradley RS, Lewis RM, Aplin JD, Johnstone ED. Whole organ vascular casting and microCT examination of the human placental vascular tree reveals novel alterations associated with pregnancy disease. Sci Rep Springer US, 2017;7:4144.

Kiessling F, Greschus S, Lichy MP, Bock M, Fink C, Vosseler S, Moll J, Mueller MM, Fusenig NE, Traupe H, Semmler W. Volumetric computed tomography (VCT): a new technology for noninvasive, high-resolution monitoring of tumor angiogenesis. Nat Med 2004;10:1133–1138.

Kobayashi H, Kawamoto S, Saga T, Sato N, Hiraga A, Konishi J, Togashi K, Brechbiel MW. Micro-MR angiography of normal and intratumoral vessels in mice using dedicated intravascular MR contrast agents with high generation of polyamidoamine dendrimer core: Reference to pharmacokinetic properties of dendrimer-based MR contrast agents. J Magn Reson Imaging 2001;14:705–713.

Kourembanas S, Hannan RL, Faller D V. Oxygen tension regulates the expression of the platelet-derived growth factor-B chain gene in human endothelial cells. J Clin Invest 1990;86:670–4.

Laufer J, Johnson P, Zhang E, Treeby B, Cox B, Pedley B, Beard P. In vivo preclinical photoacoustic imaging of tumor vasculature development and therapy. J Biomed Opt 2012;17:056016.

Lee DT. Medial Axis Transformation of a Planar Shape. IEEE Trans Pattern Anal Mach Intell 1982;PAMI-4:363–369.

Leunig M, Yuan F, Menger MD, Boucher Y, Goetz AE, Messmer K, Jain RK. Angiogenesis, Microvascular Architecture, Microhemodynamics, and Interstitial Fluid Pressure during Early Growth of Human Adenocarcinoma LSI74T in SCID Mice 1. Cancer Res 1992;52:6553–6560.

Macé E, Montaldo G, Osmanski B, Cohen I, Fink M, Tanter M. Functional ultrasound imaging of the brain: theory and basic principles. IEEE Trans Ultrason Ferroelec Freq Contr 2013;60:492–506.

Montaldo G, Tanter M, Bercoff J, Benech N, Fink M. Coherent plane-wave compounding for very high frame rate ultrasonography and transient elastography. IEEE Trans Ultrason Ferroelectr Freq Control 2009;56:489–506.

Nagy JA, Chang SH, Shih SC, Dvorak AM, Dvorak HF. Heterogeneity of the tumor vasculature. Semin Thromb Hemost 2010;36:321–331.

Opacic T, Dencks S, Theek B, Piepenbrock M, Ackermann D, Rix A, Lammers T, Stickeler E, Delorme S, Schmitz G, Kiessling F. Motion model ultrasound localization microscopy for preclinical and clinical multiparametric tumor characterization. Nat Commun 2018 [cited 2018 Aug 8];9. Available from: http://www.nature.com/articles/s41467-018-03973-8

Payen T, Dizeux A, Baldini C, Le Guillou-Buffello D, Lamuraglia M, Comperat E, Lucidarme O, Bridal SL. VEGFR2-Targeted Contrast-Enhanced Ultrasound to Distinguish between Two Anti-Angiogenic Treatments. Ultrasound Med Biol 2015;41:2202–2211.

Pugh CW, Ratcliffe PJ. Regulation of angiogenesis by hypoxia: role of the HIF system. Nat Med 2003;9:677–684.

Semenza GL. Targeting HIF-1 for cancer therapy. Nat Rev Cancer 2003;3:721–732.

Shweiki D, Itin A, Soffer D, Keshet E. Vascular endothelial growth factor induced by hypoxia may mediate hypoxia-initiated angiogenesis. Nature 1992;359:843–845.

Vakoc BJ, Lanning RM, Tyrrell JA, Padera TP, Bartlett LA, Stylianopoulos T, Munn LL, Tearney GJ, Fukumura D, Jain RK, Bouma BE. Three-dimensional microscopy of the tumor microenvironment in vivo using optical frequency domain imaging. Nat Med 2009;15:1219–1223.

Weis SM, Cheresh DA. Tumor angiogenesis: molecular pathways and therapeutic targets. Nat Med 2011;17:1359–70.

Yao J, Wang L V. Photoacoustic brain imaging: from microscopic to macroscopic scales. Neurophotonics 2014;1:011003.

Yu JL, Rak JW, Carmeliet P, Nagy A, Kerbel RS, Coomber BL. Heterogeneous vascular dependence of tumor cell populations. Am J Pathol American Society for Investigative Pathology, 2001;158:1325–1334.

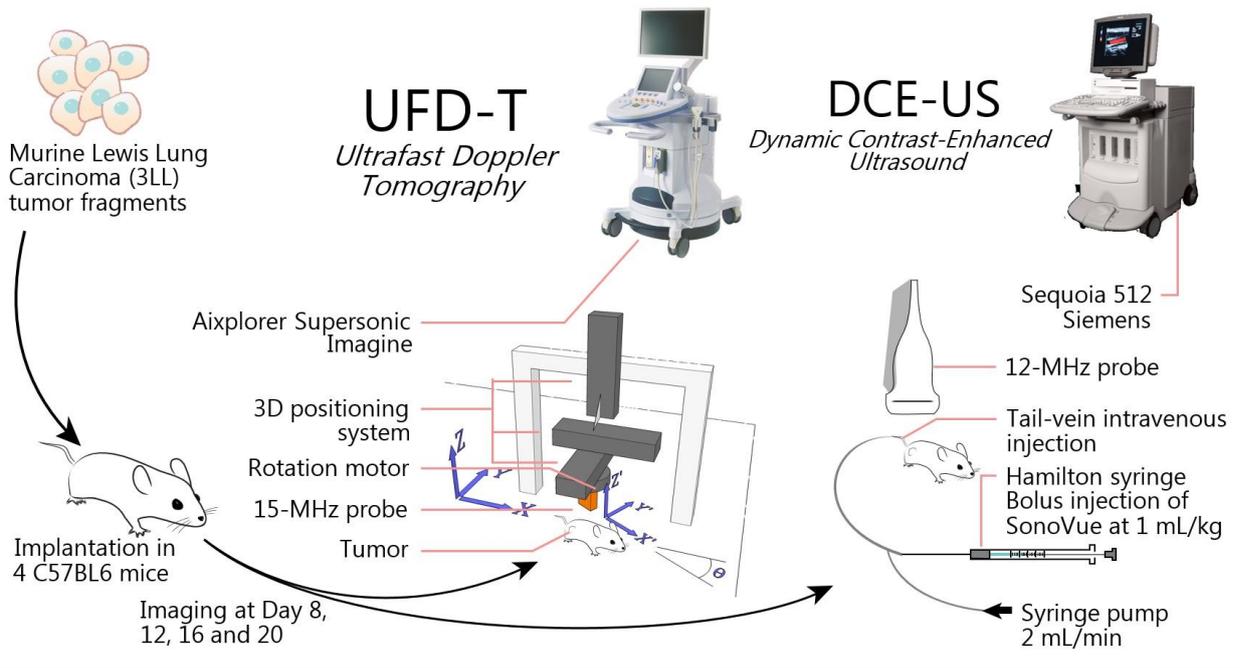

**Figure 1** *Schematic representation of the experimental protocol from tumor implantation (left), to imaging using UFD-T (middle) and DCE-US (right).*

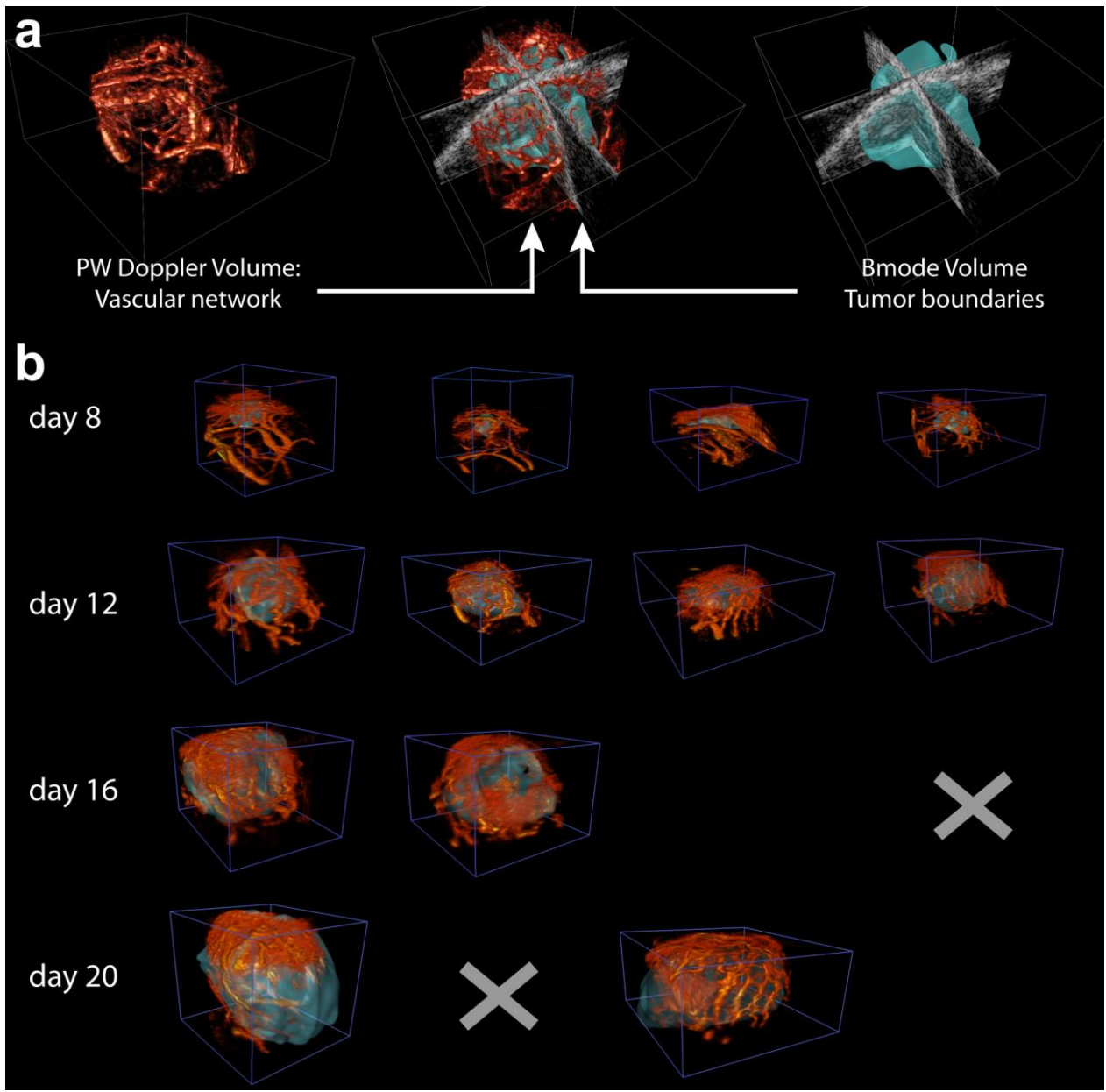

**Figure 2 *a*.** *Structural information given via UFD-T imaging: PW Doppler enables a 3D reconstruction of the tumor vascular 3D reconstruction (left image) whereas tissue images available before spatiotemporal filtering enables 3D segmentation of the tumor boundary (right image). Combined, they give a precise delineation of the vascular architecture inside and surrounding the tumor **b.** The 12 UFD-T acquisitions realized on the 4 mice of the study, represented as a Volume rendering: the transparent blue volume is the segmented contour of the tumor, and in red is the vascular network reconstructed via UFD-T. Two mice were euthanized before the end of the 20 days, and the Day 16 acquisition of Mouse 3 was not conducted. Overall, UFD-T reconstruction shows a clear peripheral organization of the tumor vascularization.*

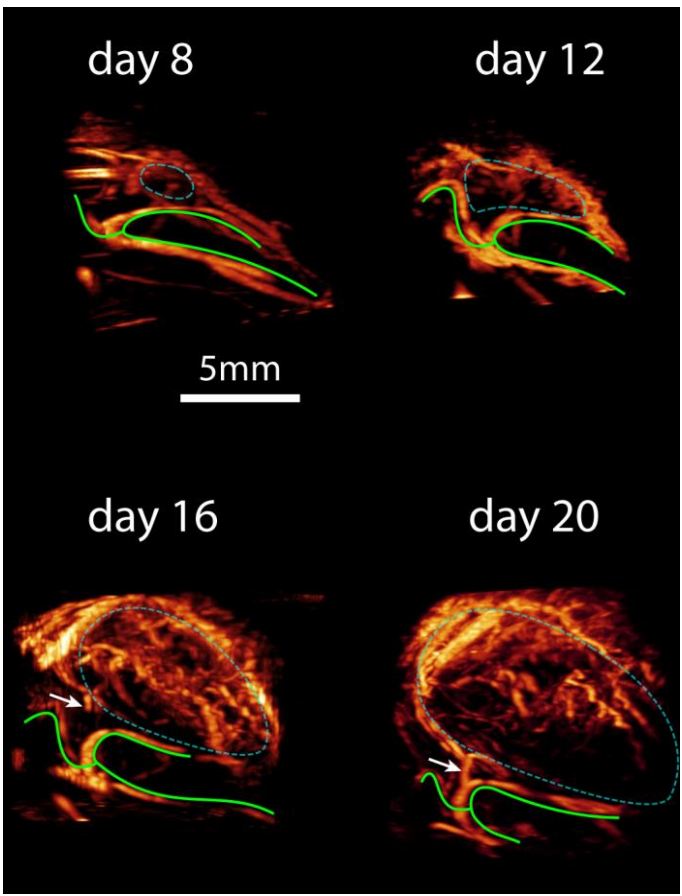

**Figure 3** *Monitoring via UFD-T of one of the four tumors on 12 days, visualized via a Maximum intensity Projection (MIP) technique. Position of the tumor is depicted as a dashed blue line. Growth of the vascular network is easily observed. Highlighted in green, a skin pre-existing large vessel , at distance from the tumor at day 8, becomes a major supply vessel for the tumor at day 20 (white arrow indicates a visible branching at day 16 and 20).*

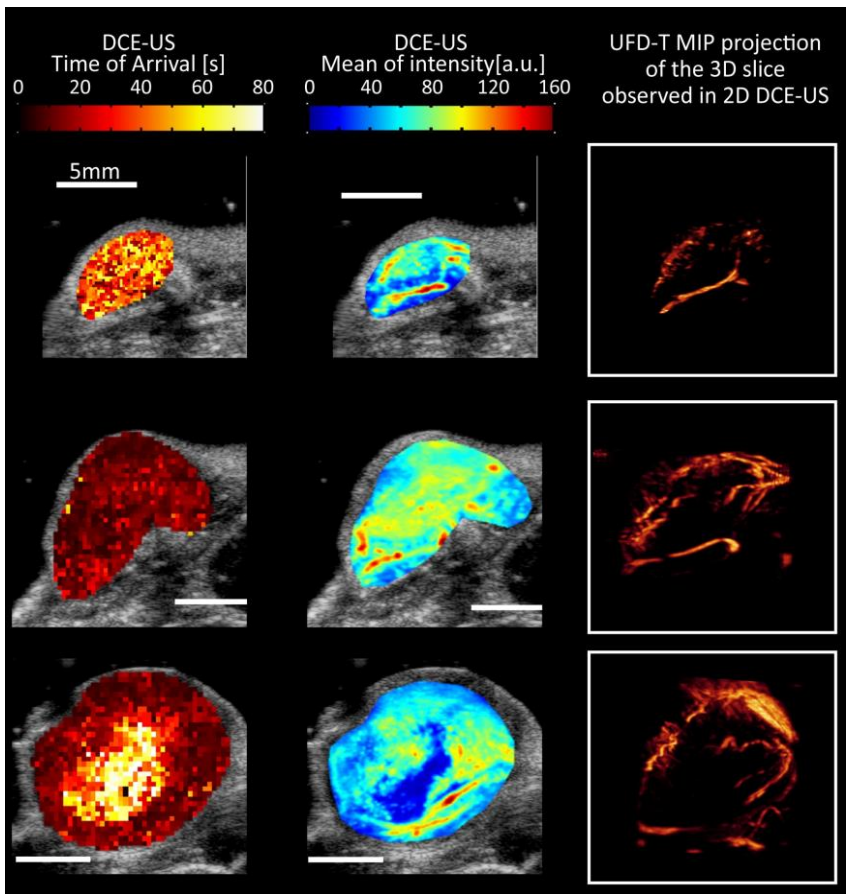

**Figure 4** *Comparison of the 2D parametric description of the tumor using DCE-US and the vascular information obtained with UFD-T in the same location (slice selected in the 3D volume) for a tumor at different stages of development (top row at Day 8, middle row at Day 12 and bottom row at Day 20). DCE-US Time of Arrival (left column) reflects the structural organization of the tumor vasculature when the Mean of Intensity (middle column) shows the level of perfusion. UFD-T Maximum Intensity Projection maps (right column) depicts the underlying vasculature which develops from a poorly ramified network to a more homogeneous tree. When the critical size is reached, the tumor core becomes very weakly perfused and the vasculature is then mainly located at the periphery.*

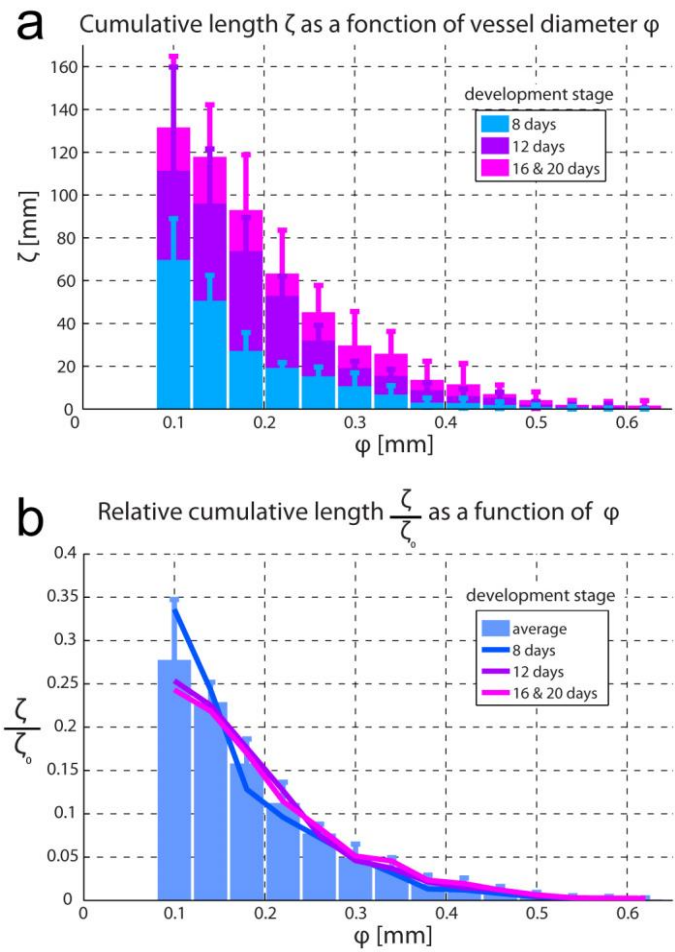

**Figure 5 *a.*** *Distribution of the blood vessels for 3 development stages in terms of cumulative length ζ plotted against the blood vessel diameters φ, for 4 tumors in each development stage (8 days, 12 days, 16 and 20 days). Each bin gathers the total cumulative length in mm of a population of vessels whose diameter is included between $\varphi \pm \epsilon$, with $\epsilon = 0.02mm$. Days 16 and 20 have been pooled to increase the number of individuals and because the difference of development at that stage was not significantly different (maximum size of the tumor is reached).* ***b.*** *Cumulative length ζ normalized by the total length of the vascular network $\zeta_0$, plotted against the same categorization of diameter. Lines represent the 3 stages of development, and the blue bars the average relative cumulative length across all tumors, whatever their development stage. For example, vessels with diameters between 0.08 and 0.16 mm represent on average ≈50% of the total length of the network at all stages of development studied.*

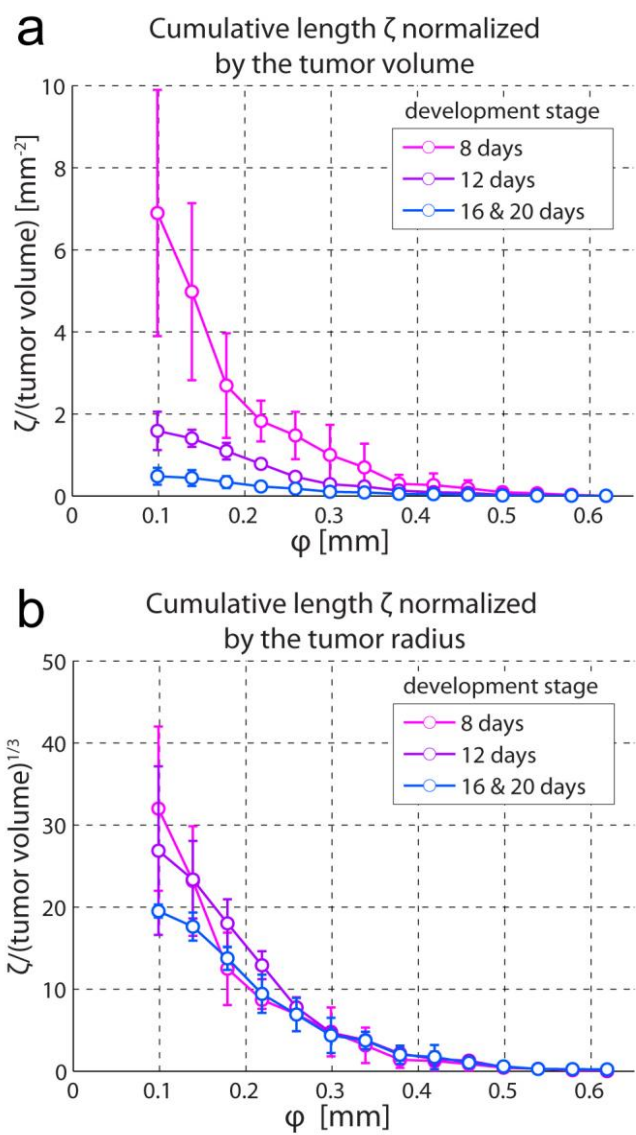

**Figure 6 a.** Cumulative length normalized by the tumor volume $\zeta/volume_{tumor}$ plotted against the vessels diameter $\varphi$. Once again each point (bins have been changed to points for the sake of readability) gathers the total cumulative length in mm of a population of vessels whose diameter is included between $\varphi \pm \epsilon$, with $\epsilon = 0.02$ mm, and day 16 and day 20 have been gathered in one group. **b.** Same graph, but the cumulative length $\zeta$ has been normalized by a scale proportional to the tumor radius, estimated as the cubic root of the tumor volume.